 \documentclass[10pt,preprint]{aastex}  %compact preprint style (MJA)

\def\gapprox{\lower.4ex\hbox{$\;\buildrel >\over{\scriptstyle\sim}\;$}}
\def\lapprox{\lower.4ex\hbox{$\;\buildrel <\over{\scriptstyle\sim}\;$}}

\def\ref#1{\par\noindent\hangindent1cm {#1}}

\shortauthors{ASCHWANDEN AND MCTIERNAN 2010}
\shorttitle{Solar Flare Waiting Times}

\begin{document}
%{\sl  Manuscript, 2010-Jan-10}

\title{		Reconciliation of Waiting Time Statistics
		of Solar Flares Observed in Hard X-Rays		}

\author{        Markus J. Aschwanden}

\affil{         Lockheed Martin Advanced Technology Center,
                Solar \& Astrophysics Laboratory,
                Org. ADBS, Bldg.252,
                3251 Hanover St.,
                Palo Alto, CA 94304, USA;
                e-mail: aschwanden@lmsal.com}

\author{	James M. McTiernan}

\affil{		Space Sciences Laboratory,
		University of California,
		Berkeley, CA 94720, USA;
		e-mail: jimm@ssl.berkeley.edu}

\begin{abstract}
We study the waiting time distributions of solar flares observed
in hard X-rays with ISEE-3/ICE, HXRBS/SMM, WATCH/GRANAT, BATSE/CGRO,
and RHESSI. Although discordant results and interpretations have been
published earlier, based on relatively small ranges ($< 2$ decades) 
of waiting times, we find that all observed distributions,
spanning over 6 decades of waiting times ($\Delta t \approx 10^{-3}-
10^3$ hrs), can be reconciled with a single distribution function,
$N(\Delta t) \propto \lambda_0 (1 + \lambda_0 \Delta t)^{-2}$, which has
a powerlaw slope of $p \approx 2.0$ at large waiting times
($\Delta t \approx 1-1000$ hrs) and flattens out at short waiting times 
$\Delta t \lapprox \Delta t_0 = 1/\lambda_0$.
We find a consistent breakpoint at $\Delta t_0 = 1/\lambda_0 = 0.80\pm0.14$ 
hours from the WATCH, HXRBS, BATSE, and RHESSI data. 
The distribution of waiting times is invariant for
sampling with different flux thresholds, while the mean waiting time 
scales reciprocically with the number of detected events, $\Delta t_0
\propto 1/n_{det}$. 
This waiting time distribution can be modeled with a nonstationary 
Poisson process with a flare rate $\lambda=1/\Delta t$ that varies as 
$f(\lambda) \propto \lambda^{-1} \exp{-(\lambda/\lambda_0)}$.
This flare rate distribution represents a highly intermittent flaring 
productivity in short clusters with high flare rates, separated by 
quiescent intervals with very low flare rates. 

\end{abstract}


\section{       	INTRODUCTION  			}

You might drive a car in a foreign city and have to stop at many red
traffic lights. From the statistics of waiting times you probably can
quickly figure out which signals operate independently and which ones
operate the smart way with inductive-loop traffic detectors in a
so-called {\sl demand-actuated mode}. Thus, statistics of waiting times
bears crucial information how a system works, either having independent 
elements that act randomly, or consisting of elements with long-range 
connections that enable coupling and synchronization. In geophysics,
aftershocks have been found to exhibit different waiting time statistics
(Omori's law) than the main shocks of earthquakes (e.g., Omori 1895;
Bak et al.~2002; Saichev and Sornette 2006). In magnetospheric physics,
waiting time statistics is used to identify Poisson 
random processes, self-organized criticality, intermittent turbulence, 
finite system size effects, or clusterization, such as in 
auroral emission (Chapman et al.~1998, 2001), the auroral electron jet (AE)
index (Lepreti et al.~2004; Boffetta et al.~1999), or in substorms at the
Earth's magnetotail (Borovsky et al.~1993; Freeman and Morley 2004).
Waiting time statistics is studied intensely in solar physics, where
most flares are found to be produced by a Poissonian random process, but
there are also so-called {\sl sympathetic flares} that have a causal connection
or trigger each other (e.g., Simnett 1974; Gergely and Erickson 1975;
Fritzova-Svestkova et al.~1976; Pearce and Harrison 1990; Bumba and Klvana
1993; Biesecker and Thompson 2000; Moon et al.~2002). Waiting time
statistics of solar flares was studied in hard X-rays (Pearce et al.~1993;
Biesecker 1994; Crosby 1996; Wheatland et al.~1998; Wheatland and Eddy 1998),
in soft X-rays (Wheatland 2000a; Boffetta et al.~1999; Lepreti et al.~2001;
Wheatland 2001; Wheatland and Litvinenko 2001; Wheatland 2006;
Wheatland and Craig 2006; Moon et al.~2001; Grigolini et al.~2002),
for coronal mass ejections (CMEs) (Wheatland 2003; Yeh et al.~2005;
Moon et al.~2003), for solar radio bursts (Eastwood et al.~2010), and for
the solar wind (Veltri 1999; Podesta et al.~2006a,b, 2007, Hnat et al.~2007;
Freeman et al.~2000; Chou 2001; Watkins et al.~2001a,b, 2002; Gabriel and
Patrick 2003; Bristow 2008; Greco et al. 2009a,b). In astrophysics,
waiting time distributions have been studied for flare stars
(Arzner and Guedel 2004) as well as for black-hole candidates, such as
Cygnus X-1 (Negoro et al.~1995). An extensive review on waiting time statistics
can be found in chapter 5 of Aschwanden (2010).

In this study we focus on waiting time distributions of solar flares
detected in hard X-rays. The most comprehensive sampling of solar flare 
waiting times was gathered in soft X-rays so far, using a 25-year catalog 
of GOES flares (Wheatland 2000a; Boffetta et al.~1999; Lepreti et al.~2001),
but three different interpretations were proposed, using the very same data: 
(i) a not-stationary
(time-dependent) Poisson process (Wheatland 2000a), (ii) a shell-model
of MHD turbulence (Boffetta et al.~1999), or (iii) a L\'evy flight
model of self-similar processes with some memory (Lepreti et al.~2001).
All three interpretations can produce a powerlaw-like distribution of
waiting times. On the other side, self-organized criticality models
(Bak et al. 1987, 1988; Lu and Hamilton 1991; Charbonneau et al.~2000) predict
a Poissonian random process, which has an exponential distribution of
waiting times for a stationary (constant) flare rate, but can produce
powerlaw-like waiting time distributions with a slope of $p \lapprox 3$
for nonstationary variations of the flare rate (Wheatland and Litvinenko
2002). Therefore, the finding of a powerlaw-like distribution of waiting
times of solar flares has ambiguous interpretations. The situation in
solar flare hard X-rays is very discordant: Pearce et al.~(1993)
and Crosby (1994) report powerlaw distributions of waiting times with
a very flat slope of $p \approx 0.75$, Biesecker (1994) reports a
near-exponential distribution after correcting for orbit effects,
while Wheatland et al.~(1998) finds a double-hump distribution with an
overabundance of short waiting times ($\Delta t \approx 10$ s - 10 min)
compared with longer waiting times ($\Delta t \approx 10-1000$ min), but
is not able to reproduce the observations with a nonstationary Poisson
process. In this study we analyze flare catalogs from HXRBS/SMM, BATSE/CGRO 
and RHESSI and are able to model all observed hard X-ray waiting time
distributions with a unified model in terms of a nonstationary Poisson 
process in the limit of high intermittency. We resolve also the 
discrepancy between exponential and powerlaw-like waiting time distributions, 
in terms of selected fitting ranges.

\section{	Bayesian Waiting Time Statistics		}
 
The waiting time distribution $P(\Delta t)$ for a {\sl Poissonian 
random process} can be approximated with an exponential distribution,
\begin{equation}
	P(\Delta t) = \lambda e^{-\lambda \Delta t} \ ,
\end{equation}
where $\lambda$ is the mean event occurrence rate, with the probability
$\int P(\Delta t) d\Delta t=1$ normalized to unity. If the average event rate 
$\lambda$ is time-independent, we call it a {\sl stationary Poisson process}. 
If the average rate varies
with time, the waiting time distribution reflects a superposition of multiple
exponential distributions with different e-folding time scales, and may even
ressemble a powerlaw-like distribution. The statistics of such {\sl inhomogeneous}
or {\sl nonstationary Poisson processes} can be characterized with 
{\sl Bayesian statistics}. 

A nonstationary Poisson process may be subdivided into time intervals where
the occurrence rate is constant, within the fluctuations expected from Poisson
statistics, so it consists of piecewise stationary processes, e.g., 
\begin{equation}
	P(\Delta t) = \left\{ \begin{array}{ll}
	  \lambda_1 e^{-\lambda_1 \Delta t} & {\rm for} \ t_1 \le t \le t_2 \\
	  \lambda_2 e^{-\lambda_2 \Delta t} & {\rm for} \ t_2 \le t \le t_3 \\
	...............		        & .......			\\
	\lambda_n e^{-\lambda_n \Delta t} & {\rm for} \ t_n \le t \le t_{n+1}
	\end{array} \right.
\end{equation}
where the occurrence rate $\lambda_i$ is stationary during a time interval 
$[t_i, t_{i+1}]$, but has different values in subsequent time intervals. The 
time intervals $[t_i, t_{i+1}]$ where the occurrence rate is stationary are 
called {\sl Bayesian blocks}, and we talk about {\sl Bayesian statistics} 
(e.g., Scargle 1998). The variation of the occurrence rates $\lambda_1, 
\lambda_2, ..., \lambda_n$ can be defined with a new time-dependent function 
$\lambda(t)$ and the probability function of waiting times becomes 
itself a function of time (e.g., Cox \& Isham 1980; Wheatland et al.~1998),
\begin{equation}
	P(t, \Delta t) = \lambda(t+\Delta t) 
	\exp{\left[-\int_t^{t+\Delta t} \lambda(t') dt' \right]} \ .
\end{equation}
If observations of a nonstationary Poisson process are made for the time 
interval $[0, T]$, then the distribution of waiting times for that time 
interval will be, weighted by the number of events $\lambda(t) dt$ in
each time interval $(t, t+dt)$,   
\begin{equation}
	P(\Delta t) = {1 \over N} \int_0^T \lambda(t) \  P(t, \Delta t) dt \ , 
\end{equation}
where the rate is zero after the time interval $t>T$, i.e., $\lambda(t>T)=0$,
and $N=\int_0^T \lambda(t) dt$.  If the 
rate is slowly varying, so that it can be subdivided into piecewise stationary
Poisson processes (into Bayesian blocks), then the distribution of waiting 
times will be  
\begin{equation}
	P(\Delta t) \approx 
	\sum_i \varphi(\lambda_i) \lambda_i e^{-\lambda_i \Delta t} \ ,
\end{equation}
where 
\begin{equation}
	\varphi(\lambda_i)={\lambda_i t_i \over \sum_j \lambda_j t_j} \ ,
\end{equation}
is the fraction of events associated with a given rate $\lambda_i$ 
in the (piecewise) time interval $t_i$ (or Bayesian block) over which 
the constant rate $\lambda_i$ is observed. If we make the transition from
the summation over discrete time intervals $t_i$ (Eqs.~5 and 6) to a 
continuous integral function over the time interval $[0<t<T]$, we obtain,
\begin{equation}
	P(\Delta t) =
	{\int_0^T \lambda(t)^2 e^{-\lambda(t) \Delta t} dt
	\over \int_0^T \lambda(t) dt} \ .
\end{equation}

When the occurrence rate $\lambda(t)$ is not a simple function,
the integral Eq.~(7) becomes untractable, in
which case it is more suitable to substitute the integration variable $t$
with the variable $\lambda$. Defining $f(\lambda)=(1/T) dt(\lambda)/d\lambda$ 
as the fraction of time that the flaring rate is in the range $(\lambda,
\lambda+d\lambda)$, or $f(\lambda) d\lambda = dt/T$, 
we can express Eq.~(7) as an integral of
the variable $\lambda$,
\begin{equation}
	P(\Delta t) =
	{\int_0^{\infty} 
	f(\lambda) \lambda^2 e^{-\lambda \Delta t} d\lambda
	\over \int_0^{\infty} \lambda f(\lambda)\ d\lambda} \ ,
\end{equation}
where the denominator $\lambda_0=\int_0^\infty \lambda f(\lambda) d\lambda$
is the mean rate of flaring.

\medskip
Let us make some examples. In Fig.~1 we show five cases: (1) a stationary 
Poisson process with a constant rate $\lambda_0$; (2) a two-step process with 
two different occurrence rates $\lambda_1$ and $\lambda_2$; 
(3) a nonstationary Poisson process with piece-wise linearly increasing 
occurrence rates $\lambda(t)=\lambda_0 (t/T)$, varying like a triangular 
function for each cycle, (4) piece-wise exponential functions, 
and (5) piece-wise exponential function steepened by a reciprocal factor. 
For each case we show the time-dependent occurrence rate $\lambda(t)$ and 
the resulting probability distribution $P(\Delta t)$ of 
events. We see that a stationary Poisson process produces an exponential 
waiting time distribution, while nonstationary Poisson processes with a 
discrete number of occurrence rates $\lambda_i$ produce a superposition of 
exponential distributions, and continuous occurrence rate functions
$\lambda(t)$ generate powerlaw-like waiting time distributions 
at the upper end. 

We can calculate the analytical functions for the waiting time distributions 
for these five cases. The first case is simply an exponential function as given
in Eq.~(1) because of the constant rate $\lambda(t)=\lambda_0$,
\begin{equation}
	P(\Delta t) = \lambda_0 e^{-\lambda_0 \Delta t} \ .
\end{equation}

The second case follows from Eq.~(5) and yields
\begin{equation}
	P(\Delta t) = 
		{1 \over 10} \lambda_1 e^{-\lambda_1 \Delta t} + 
	        {9 \over 10} \lambda_2 e^{-\lambda_2 \Delta t} \ .
\end{equation}

The third case can be integrated with Eq.~(7). The time-dependent flare
rate grows linearly with time to a maximum rate of $\lambda_m=2 \lambda_0$
over a time interval $T$, with a mean rate of $\lambda_0$,
\begin{equation}
	\lambda(t) = \lambda_m \left( { t \over T} \right) =
	             2 \lambda_0 \left( { t \over T} \right)
\end{equation}
Defining a constant $a=-\lambda_m(\Delta t/T)$ the integral of Eq.~7 yields
$P(\Delta t) = (2 \lambda_m/T^3) \int_0^T t^2 e^{at} dt$. The integral
$\int x^2 e^{ax} dx= e^{ax}(x^2/a-2x/a^2+2/a^3)$ can be obtained from an
integral table. The analytical function of the waiting time distribution
for a linearly increasing occurrence rate is then
\begin{equation}
	P(\Delta t) = 2 \lambda_m 
	\left[{2 \over (\lambda_m \Delta t)^3} -
		e^{-\lambda_0 \Delta t}
		\left({1 \over (\lambda_m \Delta t)} 
		     +{2 \over (\lambda_m \Delta t)^2} 
		     +{2 \over (\lambda_m \Delta t)^3}\right)\right] \ ,
\end{equation}
which is a flat distribution for small waiting times and approaches a 
powerlaw function with a slope of $p=3$ at large waiting times, i.e., 
$P(\Delta t) \propto \Delta t^{-3}$ (Fig.~1, third case).
The distribution is the same for a single linear ramp or a cyclic
triangular variation, because the total time spent at each rate 
$[\lambda, \lambda+d\lambda]$ is the same.
		
The fourth case, which mimics the solar cycle,
has an exponentially growing (or decaying) occurrence rate, i.e., 
\begin{equation}
	f(\lambda) =\left( { 1 \over \lambda_0} \right)
	\exp{\left(-{\lambda \over \lambda_0} \right)} \ ,
\end{equation}
defined in the range of 
$[0 < \lambda < \infty]$, and has a mean of $\lambda_0$. The waiting time 
distribution can therefore be written with Eq.~(8) as
\begin{equation}
	P(\Delta t) = \int_0^{\infty}
	\left({ \lambda \over \lambda_0}\right)^2 
	\exp{\left(-{\lambda \over \lambda_0} [1 + \lambda_0 \Delta t]\right)}
	d\lambda \ ,
\end{equation}
which corresponds to the integral $\int_0^{\infty} x^2 e^{ax} dx = -2/a^3$
using $a=-(1+\lambda_0 \Delta t)/\lambda_0$ and thus has the solution
$P(\Delta t) = -2/(a^3 \lambda_0^2)$, i.e.,
\begin{equation}
	P(\Delta t) = {2 \lambda_0 \over (1 + \lambda_0 \Delta t)^3} \ .
\end{equation}
For very large waiting times $(\Delta t \gg 1/\lambda_0)$ it approaches
the powerlaw limit $P(\Delta t) \approx (2/\lambda_0^2) \Delta t^{-3}$
(see Fig.~1 fourth panel). 

The fifth case has an exponentially growing occurrence rate, multiplied
with a reciprocal factor, i.e., 
\begin{equation}
	f(\lambda) = \lambda^{-1} \  
	\exp{\left(-{\lambda \over \lambda_0} \right)} \ ,
\end{equation}
and fulfills the normalization $\int_0^\infty \lambda f(\lambda) d\lambda
= \lambda_0$. The waiting time distribution can then be written with Eq.~(8) as
\begin{equation}
	P(\Delta t) = \int_0^{\infty}
	\left({\lambda \over \lambda_0}\right)  \ 
	\exp{\left(-{\lambda \over \lambda_0} [1 + \lambda_0 \Delta t]\right)}
	d\lambda \ ,
\end{equation}
which, with defining $a=-(1+\lambda_0 \Delta t)/\lambda_0$,
corresponds to the integral $\int x e^{ax} dx = (e^{ax}/a^2) (ax-1)$
and has the limit  $\int_0^{\infty} x e^{ax} dx = 1/a^2$,
yielding the solution $P(\Delta t) = 1/(a^2 \lambda_0^2)$, i.e.,
\begin{equation}
	P(\Delta t) = {\lambda_0 \over (1 + \lambda_0 \Delta t)^2} \ .
\end{equation}
For very large waiting times $(\Delta t \gg 1/\lambda_0)$ it approaches
the powerlaw limit $P(\Delta t) \approx (1/\lambda_0^2) \Delta t^{-2}$
(see Fig.~1 bottom panel). 

Thus we learn from the last four examples
that most continuously changing occurrence rates produce powerlaw-like
waiting time distributions with slopes of $p \lapprox 2, ..., 3$ at large
waiting times, despite of the intrinsic exponential distribution that
is characteristic to stationary Poisson processes. If the variability of
the flare rate is gradual (third and fourth case in Fig.~1),
the powerlaw slope of the waiting time distribution is close to 
$p \lapprox 3$. However, if the variability of the flare rate 
shows spikes like $\delta$-functions (Fig.~1, bottom), which is highly
intermittent with short clusters of flares, the distribution of waiting
times has a slope closer to $p \approx 2$. This phenomenon is also
called clusterization and has analogs in earthquake statistics, where
aftershocks appear in clusters after a main shock (Omori's law).
Thus the powerlaw slope of waiting times contains essential information 
whether the flare rate varies gradually or in form of intermittent
clusters.

\section{		DATA ANALYSIS 			}

We present an analysis of the waiting time distribution of solar flares
observed in hard X-rays, using flare catalogs obtained from the 
{\sl Ramaty High Energy Solar Spectroscopic Imager (RHESSI)},
the {\sl Compton Gamma Ray Observatory (CGRO)}, and
the {\sl Solar Maximum Mission (SMM)}, and model also previously published
data sets from SMM (Pearce et al.~1993), GRANAT (Crosby 1996), and
ISEE-3 (Wheatland 1998). 

\subsection{	RHESSI Waiting Time Analysis 		}

RHESSI (Lin et al.~2002) was launched on February 5, 2000 and still
operates at the time of writing, for a continuous period of 8 years.
The circular spacecraft orbit has a mean period of 96 min (1.6 hrs), a
mean altitude of 600 km, and an inclination of $38^\circ$ with respect
to the equator, so the observing mode is interrupted by gaps
corresponding to the spacecraft night (with a duration of about 35
percent of the orbit), as well as some other data gaps when flying
through the {\sl South Atlantic Anomaly (SAA)}. These data gaps
introduce some systematic clustering of waiting times around 
an orbital period ($\approx 1.6$ hrs).

We are using the official RHESSI flare list (which can be downloaded
from the RHESSI webpage {\sl
http://hesperia.gsfc.nasa.gov/rhessidatacenter/}). We are using data
from the first six years of the mission (from February 12, 2002, to 
December 31, 2007), when a uniform threshold for flare detection was
applied, while a lower threshold was applied after this period.
We include all flare events that have a solar origin, in the sense
that they could be imaged in the energy range of 12 to
25 keV. The complete event catalog (sampled up to Feb 1, 2010)
contains $n=52,014$ events, of which 12,379 have been confirmed to be
solar. The number of confirmed solar flare events we are using from the 
first six years includes 11,594 events, which corresponds to a mean 
flare rate of $<\lambda> \approx 5.5$ events per day during 2002-2007. 

A time series of the
daily RHESSI flare rate is shown in Fig.2 (top panel). The mean annual
flare rate clearly drops systematically from 2002 to 2008, but the
daily fluctuations are larger than the slowly-varying mean trend. We
calculate the waiting times simply from the time difference between
two subsequent flare events,
\begin{equation}
	\Delta t_i = t_i - i_{i-1} \ .
\end{equation}
If we plot the waiting time distribution on a log-log scale
(Fig.~2 bottom right), we see that the waiting time distribution
$N(\Delta t) = n P(\Delta t)$ can approximately be fitted with a 
powerlaw function, i.e., $N(\Delta t) \propto \Delta t^{-p}$,
with a powerlaw slope of $p \approx 2.0$. However,
there is some clustering of events above one orbital period of 
$\Delta t \gapprox t_{orbit} \approx 1.6$ hrs, as well as at the 
second to fourth harmonics of the orbital
period (Fig.~1, bottom left), which are caused
for several reasons. A first reason is that events that start at spacecraft
night cannot be detected until the spacecraft comes out of the night part of 
the orbit, causing some clustering after one full orbital period. 
A second reason is that large events that extend over more than one 
full orbital period are double-counted in each subsequent orbit. These
instrumental biases have been modeled with Monte-Carlo simulations in previous
waiting time studies (e.g., Biesecker 1994), but instead of correcting for
these erroneous waiting times, we will just exclude the range of
$\Delta t \approx (0.5-2) \times t_{orbit}$ ($\approx$0.4-2.4 hrs) in the 
fits of waiting time distributions. 

Thus our first result is that the waiting time distribution of RHESSI flares
is approximately consistent with a powerlaw distribution in the time range of 
$\Delta t \approx 1-1000$ hrs, with a powerlaw slope of $p \approx 2.0$.
For a nonstationary Poisson process we expect powerlaw slopes in the
range of $p \approx 2.0, ..., 3.0$ (Fig.~1), so the measured 
distribution is close to what the theory predicts for very intermittently
varying flare rates (Fig.~1, bottom). The high degree of intermittency 
is indeed clearly recognizable in the data (Fig.~2, top panel), where
the short-term fluctuations are much faster than the long-term trend.
One might subdivide the flare activity into two states, i.e., 
{\sl ``quiescent''} and {\sl ``flare-active''} states, similar to the
quiescent periods (soft state) and active periods (hard state) of 
pulses from accretion disk objects and black-hole candidates
(e.g., Cygnus X-1; Mineshige et al.~1994). We indicate quiescent periods
with waiting times $\Delta t \ge 5$ hrs in Fig.~2 (top panel), where it 
can be seen that they occur in every month through all 8 years, but more
frequently during the (extended) solar minimum (2006-2008). 
Thus, flare-active periods are very intermittent and not lasting 
contiguously over extended periods of time. The situation is also similar 
to earthquakes, where aftershocks appear clustered in time intervals 
after larger main shocks (Omori's law).

In a next step we investigate the effect of flux thresholds in the event
definition on the distribution of waiting times, an issue that has been
raised in previous studies (e.g., Buchlin et al.~2005; Hamon et al.~2002).
Hamon et al.~(2002) finds for the Olami-Feder-Christensen model (Olami
et al.~1992), which is a cellular automaton model for systems with 
self-organized criticality, that event definitions without any threshold
lead to stretched exponential waiting time distributions, while 
threshold-selected events produce an excess of longer waiting times.
We investigate this problem simply by applying various flux thresholds
to the RHESSI flare catalog, e.g., $P=10, 100, 300, 1000, 3000$ cts s$^{-1}$,
and by re-sampling the waiting times for events above these flux thresholds.
The corresponding six waiting time distributions are shown in Fig.~3,
which contain $n=11,594$ events detected without a threshold, 
and $n_T=9596$, 2038, 781, 271, and 108 events for the thresholded subsets.
The waiting time distributions clearly show an increasing excess of longer 
waiting times with progressively higher thresholds, which is expected due 
to the filtering
out of shorter time intervals between flares with weaker fluxes below the
threshold. Based on the reduction in the number $n$ of events as a function
of the flux threshold, we can make a prediction how the mean waiting time
interval increases with increasing threshold, namely a reciprocal relationship,
since the total duration $T$ of waiting times is constant,
\begin{equation}
	T = \sum_i^n \Delta t_i = <n> <\Delta t> 
	  = <n_T> <\Delta t>_T \ .
\end{equation}
Thus, from the number of events $n_T$ detected above a selected 
threshold we can predict the mean waiting time,
\begin{equation}
	 <\Delta t>_T = { <n> \over <n_T> } <\Delta t> \ .
\end{equation}
Using the full set of data with $<n>=11,594$ events and a mean waiting
time of $<\Delta t>=0.71$ hrs (Fig.~3, top left), we can predict the
distributions and average waiting times for the thresholded subsets,
based on their detected numbers $n_T$ using Eq.~(21):
$<\Delta t>_T = 0.9, 4.0, 10.5, 30.4, 76.3$ hrs. We fit our theoretical
model of the waiting time distribution (Eq.~18) of a nonstationary 
Poisson process and predict the distributions for the threshold datasets,
using the scaling of the predicted average waiting times $<\Delta t>_T$.
The predicted distributions (thick curves in Fig.~3) match the observed
distributions of thresholded waiting times (histograms with error bars
in Fig.~3) quite accurately, and thus demonstrates how the waiting time 
distribution changes in a predictable way when flux thresholds are used 
in the event selection. 

Regarding the mean flare waiting time we have to distinguish between the 
theoretical model value $\Delta t_0$ and the mean detected interval
$<\Delta t>^{obs}$. The theoretical value is calculated based on a complete
distribution in the range of $\Delta t=[0, \infty]$. For RHESSI
data we obtain a value of $\Delta t_0=0.71$ hrs. The observational value,
which is about the total observing time span ($\approx 5.8$ yrs) multiplied
with the spacecraft duty cycle, which is about $q=0.65$ for RHESSI (based on
spacecraft nights with a length of 32-38 minutes), and divided by the
number of observed events $n$. So, we obtain a mean time interval of 
$<\Delta t>^{obs} = T q / n = 5.8 \times 365 \times 24 \times 0.65
/ 11,594 \approx 3.0$ hrs. This value is about a factor of 4 longer
than the theoretical value. This discrepancy results from either
missing short waiting times predicted by the model (since
the observed distribution falls off at waiting times of $\Delta t \lapprox
0.02$ hrs, i.e., $\approx 1$ minute), or because the model overpredicts
the maximum flare rate, and thus needs to be limited at the maximum
flare rate or lower cutoff of waiting times. Neverthelss, whatever 
lower cutoff of observed waiting times or what flux threshold is used, 
the theoretical value $\Delta t_0$ always indicates where the breaking 
point is in the waiting time distribution, between the powerlaw part 
and the rollover at short waiting times. 

\subsection{	BATSE/CGRO Waiting Time Analysis 		}

The Compton Gamma Ray Observatory (CGRO) was launched on April 5, 1991, and
de-orbited on June 4, 2000. The CGRO spacecraft had an orbit with a period
of $t_{orbit} \approx 1.5$ hrs also, and thus is subject to similar data gaps 
as we discussed for RHESSI, which causes a peak in the waiting time 
distribution around $\Delta t \approx (0.5-1.1) t_{orbit}$. 

We use the solar flare catalog from the {\sl Burst and Transient Source
Experiment (BATSE)} onboard CGRO, which is accessible at the NASA/GSFC
homepage {\sl http://umbra.nascom.nasa.gov/batse/}. Here we use a subset of
the flare catalog obtained during the maximum of the solar cycle between
April 19, 1991 and November 12, 1993, containing 4113 events during a span of
1.75 years, yielding a mean event rate of $<\lambda> \approx 6.4$ events per day.
BATSE has 8 detector modules, each one consisting of an uncollimated, shielded
NaI scintillation crystal with an area of 2025 cm$^{2}$, sensitive in the
energy range of 25 keV-1.9 MeV (Fishman et al. 1989). 

The waiting time distribution obtained from BATSE/CGRO is shown in Fig.~4
(middle right), which ranges from $\Delta t \approx 0.01$ hrs up to 
$\approx 200$ hrs. We fit the same theoretical model of the waiting
time distribution (Eq.~18) as for RHESSI, and the powerlaw slope of
$p=2$ fits equally well. Since the obtained mean flare
rate ($\Delta t_0 = 0.92$ hrs) is similar to RHESSI ($\Delta t_0=0.71$ hrs),
the thresholds for selected flare events seems to be compatible.

\subsection{	HXRBS/SMM Waiting Time Analysis 		}

While RHESSI observed during the solar cycles \#23+24, CGRO observed the 
previous cycles \#22+23, and the Solar Maximum Mission (SMM) the previous ones
\#21+22, none of them overlapping with each other. SMM was lauched on
February 14, 1980 and lasted until orbit decay on December 2, 1989.
The {\sl Hard X-Ray Burst Spectrometer (HXRBS)} (Orwig et al.~1980) is
sensitive in the range of 20-260 keV and has a detector area of 71 cm$^{-2}$.
The orbit of SMM had initially a height of 524 km and an inclination of
$18.6^\circ$, which causes similar data gaps as RHESSI and CGRO.
HXRBS recorded a total of 12,772 flares above energies of $\gapprox 20$ keV
during a life span of 9.8 years, so the average flare rate is
$<\lambda> \approx 3.6$ flares per day, which is about half that of BATSE
and slightly less than RHESSI, which results as a combination of different
instrument sensitivities, energy ranges, and flux thresholds. 

The waiting time distribution obtained from HXRBS/SMM is shown in Fig.~4
(top right), which has a similar range of $\Delta t \approx 0.01-500$ hrs
as BATSE and RHESSI. We fit the same waiting time distribution (Eq.~18) 
with a powerlaw slope of $p = 2$ and find similar best-fit value for
the average waiting times, i.e., $\Delta t_0=1.08$ hrs, so the sensitivity
and threshold is similar.  Also the orbital effects modify the observed
distributions exactly the same way. Thus we can interpret the distributions
of waiting times for HXRBS in the same way as for BATSE and RHESSI, in terms
of a nonstationary Poisson process with high intermittency.

An earlier study on the waiting time distribution of HXRBS/SMM flares was
published by Pearce et al.~(1993), containing a subset of 8319 events
during the first half of the mission, 1980-1985 (Fig.~4, top left panel).
However, the published
distribution contained only waiting times in a range of $\Delta t=1-60$ min
($\approx 0.02-1.0$ hrs), and was fitted with a powerlaw function with a 
slope of
$p=0.75\pm0.1$ (Pearce et al.~1993; Fig.~5 therein). We fitted the same
intermediate range of waiting times ($\Delta t \approx 0.1-1.0$ hrs)
and found similar powerlaw slopes, i.e. $p \approx 0.72$ for HXRBS, 
$p = 0.84$ for BATSE, and $p=0.65$ for RHESSI, so all distributions seem 
to have a consistent slope in this range.
This partial powerlaw fits are indicated with a straight line in the
range of $\Delta t \approx 0.02-1.0$ hrs in all six cases shown in Fig.~4.
However, the waiting time distribution published by Pearce et al.~(1993)
extends only over 1.7 orders of magnitude, and thus does not reveal the
entire distribution we obtained over about 5 orders of magnitude
with HXRBS, BATSE, and RHESSI. If we try to fit the same waiting time
distribution (Eq.~18) to the dataset of Pearce et al.~(1993), we obtain
a similar fit but cannot constrain the powerlaw slope for longer waiting 
times.

\subsection{	WATCH/GRANAT Waiting Time Analysis 		}

A waiting time distribution of solar flares has earlier been published
using WATCH/GRANAT data (Crosby 1996), obtained from the Russian GRANAT
satellite, launched on December 1, 1989. GRANAT has a highly eccentric
orbit with a period of 96 hrs, a perigee of 2000 km, and an apogee of
200,000 km. Such a high orbit means that the spacecraft can observe the
Sun uninterrupted without Earth occultation (i.e., no spacecraft night),
which makes the waiting time distribution complete and free of data gaps.
WATCH is the {\sl Wide Angle Telescope for Cosmic Hard X-Rays}
(Lund 1981) and has a sensitive detector area of 95 cm$^2$ in the energy
range of 10-180 keV. 

Waiting time statistics was gathered during four 
time epochs: Jan-Dec 1990, April-Dec 1991, Jan-April 1992, and July 1992.
Crosby (1996) obtained a waiting time distribution for 
$n=182$ events during these epochs with waiting times in the range of 
$\Delta t \approx 0.04-15$ hrs. The distribution was found to be 
a powerlaw with a slope of $p=0.78\pm0.13$ in the range of
$\Delta t \lapprox 3$ hrs, with an exponential fall-off in the range of
$\Delta t \approx 3-15$ hrs. We reproduce the measured waiting time
distribution of Crosby (1996, Fig.~5.16 therein) in Fig.~4 (middle left
panel) and are able to reproduce the same powerlaw slope of $p=0.78$
by fitting the same range of $\Delta t \approx 0.1-2.0$ hrs as fitted
in Crosby (1996). We fitted also the model of the waiting time distribution
for a nonstationary Poisson process (Eq.~18) and find a similar mean
waiting time of $\Delta t_0=0.97$ hrs (Fig.~4), although there are
no data for waiting times longer than $\gapprox 15$ hrs. Thus, the
partial distribution measured by Crosby (1996) is fully consistent with
the more complete datasets analyzed from HXRBS/SMM, BATSE/CGRO, and
RHESSI.

\subsection{	ISEE-3/ICE Waiting Time Analysis 		}

Another earlier study on waiting times of solar flare hard X-ray bursts
was published by Wheatland et al.~(1998), which is of great interest here 
because it covers the largest range of waiting times analyzed previously
and the data are not 
affected by periodic orbital datagaps. The {\sl International
Sun-Earth/Cometary Explorer (ISEE-3/ICE)} spacecraft was inserted into a
``halo'' orbit about the libration point L1 some 240 Earth radii upstream
between the Earth and Sun. This special orbit warrants uninterrupted
observations of the Sun without orbital data gaps. ISEE-3 had a
duty cycle of 70-90\% during the first 5 yrs of observations, and was falling
to 20-50\% later on. ISEE-3 detected 6919 hard X-ray bursts during the
first 8 years of its mission, starting in August 1978. 

Wheatland et al.~(1998) used a flux selection of $> 4$ photons cm$^{-2}$
s$^{-1}$ to obtain a near-complete sampling of bursts, which reduced the
dataset to $n=3574$ events. The resulting waiting time distribution
extends over $\Delta t \approx 0.002-14$ hrs and could not be fitted with a
single nonstationary process, but rather exhibited an overabundance of 
short waiting times $\Delta t \lapprox 0.2$ hrs (Wheatland et al.~1998).
We reproduce the measured waiting time distribution of Wheatland et 
al.~(1998; Fig.~2 therein) in Fig.~4 (bottom left panel) and fit the 
combined waiting time distribution of a nonstationary Poisson process
(Eq.~18) and an exponential random distribution for the short waiting
times (Eq.~1). We find a satisfactory fit with the two waiting
time constants $\Delta t_0=0.03$ hrs and $\Delta t_1=1.84$ hrs.
The primary component of $\Delta t_0=0.03$ hrs ($\approx 2$ min) is
much shorter than for any other dataset, which seems to correspond to
a clustering of multiple hard X-ray bursts per flare, but is consistent
with a stationary random process itself. The secondary component contains
45\% of the hard X-ray bursts and has a waiting time scale of
$\Delta t_1=1.84$ hrs. This component seems to correspond to the
generic waiting time distribution we found for the other data, within
a factor of two. Since there are no waiting times longer than 20 hrs,
the powerlaw section of the distribution is not well constrained for
longer waiting times, but seems to be consistent with the other
datasets.

\section{			Conclusions 		}

We revisited three previously published waiting time distributions of
solar flares (Pearce et al.~1993; Crosby 1996; Wheatland et al.~1998)
using data sets from HXRBS/SMM, WATCH/GRANAT, ISEE-3/ICE, and analyzed
three additional datasets from HXRBS/SMM, BATSE/CGRO, and RHESSI.
While the previously published studies arrive at three different
interpretations and conclusions, we are able to reconcile all datasets
and the apparent discrepancies with a unified waiting time distribution
that corresponds to a nonstationary Poisson process in the limit of 
high intermittency. Our conclusions are the following:

\begin{enumerate}
\item{Waiting time statistics gathered over a relatively small range 
of waiting times, e.g., $\lapprox 2$ decades as published by
Pearce et al.~(1993) or Crosby (1996), does not provide sufficient
information to reveal the true functional form of the waiting time 
distribution. Fits to such partial distributions were found to be consistent 
with a powerlaw 
function with a relatively flat slope of $p\approx 0.75$ in the range of 
$\Delta t \approx 0.1-2.0$ hrs, but this powerlaw slope does not extend 
to shorter or longer waiting times, and thus is not representative for 
the overall distribution of waiting times.}

\item{Waiting times sampled over a large range of $5-6$ decades all
reveal a nonstationary waiting time distribution with a mean waiting
time of $\Delta t_0 = 0.80\pm0.14$ hrs (averaged
from WATCH/GRANAT, HXRBS/SMM, BATSE/CGRO, and RHESSI)
and an approximate powerlaw slope of $p \approx 2.0$.
This value of the powerlaw slope is consistent with a theoretical
model of highly intermittently variations of the flare rate,
in contrast to a more gradually changing flare rate that would produce
a powerlaw slope of $p \approx 3.0$. Waiting time studies with solar
flares observed in soft X-rays (GOES) reveal powerlaw slopes in the
range from $p \approx 1.4$ during the solar minimum to $p \approx 3.2$
during the solar maximum (Wheatland and Litvinenko 2002), which would
according to our model correspond to rather gradual variation of the 
flare rate with few quiescent periods during the solar maximum, but a 
high intermittency with long quiescent intervals during the solar minimum.
The flare rate essentially rarely drops to a low background level during
the solar maximum, so long waiting times are avoided and the frequency
distribution is steeper due to the lack of long waiting times.}

\item{For the dataset of ISEE-3/ICE there is an overabundance of
short waiting times with a mean of $\Delta t = 0.03$ hrs (2 min),
which seems to be associated with the detection of clusters of multiple
hard X-ray bursts per flare (Wheatland et al.~1998).}

\item{The threshold used in the detection of flare events modifies the
observed waiting time distribution in a predictable way. If a first
sample of $n_1$ flare events is detected with a low threshold and a second
sample of $n_2$ events with a high threshold, the waiting time distributions 
of the form $N(\Delta t) = (1/\Delta t_{i}) / (1 + \Delta t/\Delta t_{i})^2$ 
relate to each other with a reciprocal value for the mean waiting time, i.e.,
$\Delta t_2/\Delta t_1 = (n_1/n_2)$. This relationship could be confirmed 
for the RHESSI dataset for six different threshold levels.}
\end{enumerate}

The main observational result of this study is that the waiting time
distribution of solar flares is consistent with a nonstationary
Poisson process, with a highly fluctuating variability of the flare rate
during brief active periods, separated by quiescent periods with much
lower flaring variability. This behavior is 
similar to the quiescent periods (soft state) and active periods (hard state) 
of pulses from accretion disk objects and black-hole candidates
(Mineshige et al.~1994). The fact that solar flares are consistent
with a nonstationary Poissonian process
does not contradict interpretations in terms of   
self-organized criticality (Bak et al. 1987, 1988; Lu and Hamilton 1991; 
Charbonneau et al.~2000), except that the input rate or driver is 
intermittent. Alternative interpretations, such as intermittent turbulence, 
have also been proposed to drive solar flares (Boffetta et al.~1999; 
Lepreti et al.~2001), but they have not demonstrated a theoretical model 
that correctly predicts the waiting time distribution over a range of five
orders of magnitude, as observed here with five different spacecraft. 

\acknowledgements {\sl Acknowledgements:} 
This work is partially supported by NASA contract
NAS5-98033 of the RHESSI mission through University of California,
Berkeley (subcontract SA2241-26308PG) and NASA grant NNX08AJ18G. 
We acknowledge access to solar mission data and flare catalogs from the 
{\sl Solar Data Analysis Center(SDAC)} at the NASA Goddard Space Flight 
Center (GSFC).

%%%%%%%%%%%%%%%%%%%% REFERENCES %%%%%%%%%%%%%%%%%%%%%%%%%%%

\section*{REFERENCES}

\def\ref#1{\par\noindent\hangindent1cm {#1}}

\ref{Arzner, K. and Guedel, M. 2004, ApJ 602, 363.}
\ref{Aschwanden, M.J., 2010, {\sl Self-Organized Criticality in Astrophysics.
	Statistics of Nonlinear Processes in the Universe},
	PRAXIS Publishing Inc, Chichester, UK, and Springer, Berlin 
	(in preparation).}
\ref{Bak, P., Tang, C., and Wiesenfeld, K. 1987, Phys. Rev. Lett. 59/27, 381.}
\ref{Bak, P., Tang, C., \& Wiesenfeld, K. 1988, Phys. Rev. A 38/1, 364.}
\ref{Bak, P., Christensen, K., Danon, L., and Scanlon, T. 2002,
        Phys. Rev. lett. 88/17, 178,501.}
\ref{Biesecker, D.A. 1994, PhD Thesis, University of New Hampshire.}
\ref{Biesecker, D.A. and Thompson, B.J. 2000,
        J. Atmos. Solar-Terr. Phys. 62/16, 1449.}
\ref{Boffetta, G., Carbone, V., Giuliani, P., Veltri, P., and Vulpiani, A. 
	1999, Phys. Rev. Lett. 83/2, 4662.}
\ref{Borovsky, J.E., Nemzek, R.J., and Belian, R.D. 1993, 
	J. Geophys. Res. 98, 3807.}
\ref{Bristow, W. 2008, JGR 113, A11202, doi:10.1029/2008JA013203.}
\ref{Buchlin, E., Galtier, S., and Velli, M. 2005, AA 436, 355.}
\ref{Bumba, V. and Klvana, M. 1993, Solar Phys. 199, 45.}
\ref{Chapman, S.C., Watkins, N.W., Dendy, R.O., Helander, P.,
        and Rowlands, G. 1998, GRL 25/13, 2397.}
\ref{Chapman, S.C., Watkins, N., and Rowlands, G. 2001,
        J. Atmos. Solar-Terr. Phys. 63, 1361.}
\ref{Charbonneau, P., McIntosh, S.W., Liu, H.L., and Bogdan, T.J. 2001,
        Solar Phys. 203, 321.}
\ref{Chou, Y.P. 2001, Solar Phys. {\bf 199}, 345.}
\ref{Cox, D. and Isham, V. 1980, {\sl Point Processes}, 
	London: Chapman \& Hall.}
\ref{Crosby, N.B. 1996, PhD Thesis, University Paris VII, Meudon, Paris.}
\ref{Eastwood, J.P., Wheatland, M.S., Hudson, H.S., Krucker, S., Bale, S.D.,
        Maksimovic, M., Goetz, K. 2010, ApJ 708, L95.}
\ref{Fishman, G.J., Meegan, C.A., Wilson, R.B., Paciesas, W.S., Parnell, T.A., 
	Austin,R.W., Rehage, J.R., Matteson, et al.
	1989, {\sl ``CGRO Science Workshop''}, Proc. Gamma Ray Observatory 
	Science Workshop, (ed. W.N.Johnson), NASA Document 4-366-4-373, GSFC, 
	Greenbelt, Maryland, p.2-39 and p.3-47.}
\ref{Freeman, M.P., Watkins, N. W., and Riley, D. J. 2000, 
	Phys. Rev. E, 62, 8794.}
\ref{Freeman, M.P. and Morley, S.K. 2004, GRL 31, L12807.}
\ref{Fritzova-Svestkova, L., Chase, R.C., and Svestka, Z. 1976,
        Solar Phys. 48, 275.}
\ref{Gabriel, S.B. and Patrick, G.J. 2003, Space Sci. Rev. 107, 55.}
\ref{Gergely, T., and Erickson, W.C. 1975, Solar Phys. 42, 467.}
\ref{Greco, A., Matthaeus, W.H., Servidio, S., and Dmitruk, P. 2009a,
	Phys. Rev. E 80, 046401.}
\ref{Greco, A., Matthaeus, W.H., Servidio, S., Chuychai, P., and Dmitruk, P. 
	2009b, ApJ 691, L111.}
\ref{Grigolini, P., Leddon, D., and Scafetta, N. 2002,
        Phys. Rev. E 65, 046203.}
\ref{Hamon, D., Nicodemi,M., and Jensen,H.J., 2002, AA 387, 326.}
\ref{Hnat, B., Chapman, S.C., Kiyani, K., Rowlands, G., and Watkins, N.W. 2007,
        Geophys. Res. Lett. 34/15, CiteID L15108.}
\ref{Lepreti, F., Carbone, V., and Veltri, P. 2001, ApJ 555, L133.}
\ref{Lepreti, F., Carbone, V., Giuliani, P., Sorriso-Valvo, L., and Veltri, P.
        2004, Planet. Space Science 52, 957.}
\ref{Lin, R.P., Dennis, B.R., Hurford, G.J., Smith, D.M., Zehnder, A.,
	Harvey, P.R., Curtis, D.W., Pankow, D. et al. 2002, 
	Solar Phys. 210, 3.}
\ref{Lu, E.T. and Hamilton, R.J. 1991, ApJ 380, L89.}
\ref{Lund, N. 1981, ApJSS 75, 145.}
\ref{Mineshige, S., Ouchi,N.B., and Nishimori, H. 1994,
        Publ. Astron. Soc. Japan 46, 97.}
\ref{Moon, Y.J., Choe, G.S., Yun, H.S., and Park, Y.D. 2001,
        J. Geophys. Res. 106/A12, 29951.}
\ref{Moon, Y.J., Choe, G.S., Park, Y.D., Wang, H., Gallagher, P.T., Chae, J.C.,
        Yun, H.S., and Goode, P.R. 2002, ApJ 574, 434.}
\ref{Moon, Y.J., Choe, G.S., Wang, H., and Park, Y.D. 2003, ApJ 588, 1176.}
\ref{Negoro, H., Kitamoto, S., Takeuchi, M., and Mineshige, S. 1995,
        ApJ 452, L49.}
\ref{Olami, Z., Feder, H.J.S., and Christensen, K. 1992,
        Phys. Rev. Lett. 68/8, 1244.}
\ref{Omori, F., 1895, J. Coll. Sci. Imper. Univ. Tokyo 7, 111.}
\ref{Orwig, L.E., Frost, K.J., and Dennis, B.R. 1980, Solar Phys. 65, 25.}
\ref{Pearce, G. and Harrison, R.A. 1990, AA 228, 513.}
\ref{Pearce, G., Rowe, A.K., and Yeung, J. 1993, 
	Astrophys. Space Science 208, 99.}
\ref{Podesta, J.J., Roberts, D.A., and Goldstein, M.L. 2006a,
        J. Geophys. Res. 111/A10, CiteID A10109.}
\ref{Podesta, J.J., Roberts, D.A., and Goldstein, M.L. 2006b,
        J. Geophys. Res. 111/A9, CiteID A09105.}
\ref{Podesta, J.J., Roberts, D.A., and Goldstein, M.L. 2007, ApJ 664, 543.}
\ref{Press, W.H., Flannery, B.P., Teukolsky, S.A., and Vetterling, W.T.
	1986, {\sl Numerical recipes. The art of scientific computing},
	Cambridge University Press, Cambridge.}
\ref{Saichev, A. and Sornette,D. 2006, Phys. Rev. Lett. 97/7, id. 078501.}
\ref{Scargle, J. 1998, ApJ {\bf 504}, 405-418.}
\ref{Simnett, G.M. 1974, Solar Phys. 34, 377.}
\ref{Veltri, P. 1999, Plasma Phys. Controlled Fusion {\bf 41}, A787-A795.}
\ref{Watkins, N.W., Freeman, M.P., Chapman, S.C., and Dendy, R.O. 2001a, 
	J. Atmos. Solar-Terr Phys. 63, 1435.}
\ref{Watkins, N.W., Oughton, S., and Freeman, M.P. 2001b,
	Planet Space Sci. 49, 1233.}
\ref{Wheatland, M.S., Sturrock, P.A., and McTiernan, J.M. 1998, ApJ 509, 448.}
\ref{Wheatland, M.S. and Eddey, S.D. 1998,
        in Proc. Nobeyama Symposium, ``Solar Physics with Radio Observations'',
        (eds. Bastian, T., Gopalswamy, N., and Shibasaki, K.),
        NRO Report 479, p.357.}
\ref{Wheatland, M.S. 2000a, Astrophys. J. {\bf 536}, L109.}
\ref{Wheatland, M.S. 2001, Solar Phys. {\bf 203}, 87.}
\ref{Wheatland, M.S. and Litvinenko, Y.E. 2001, ApJ 557, 332.}
\ref{Wheatland, M.S. and Litvinenko, Y.E. 2002, Solar Phys. 211, 255.}
\ref{Wheatland, M.S. 2003, Solar Phys. 214, 361.}
\ref{Wheatland, M.S. 2006, Solar Phys. 236, 313.}
\ref{Wheatland, M.S. and Craig, I.J.D. 2006, Solar Phys. 238, 73.}
\ref{Yeh,, W.J., and Kao, Y.H. 1984, Phys. Rev. Lett. 53/16, 1590.}

\clearpage

%%%%%%%%%%%%%%%%%%%%%%%%%%%%%%%%%%%%%% FIGURES %%%%%%%%%%%%%%%%%%%%%%%%%%%%%%

\begin{figure}
\plotone{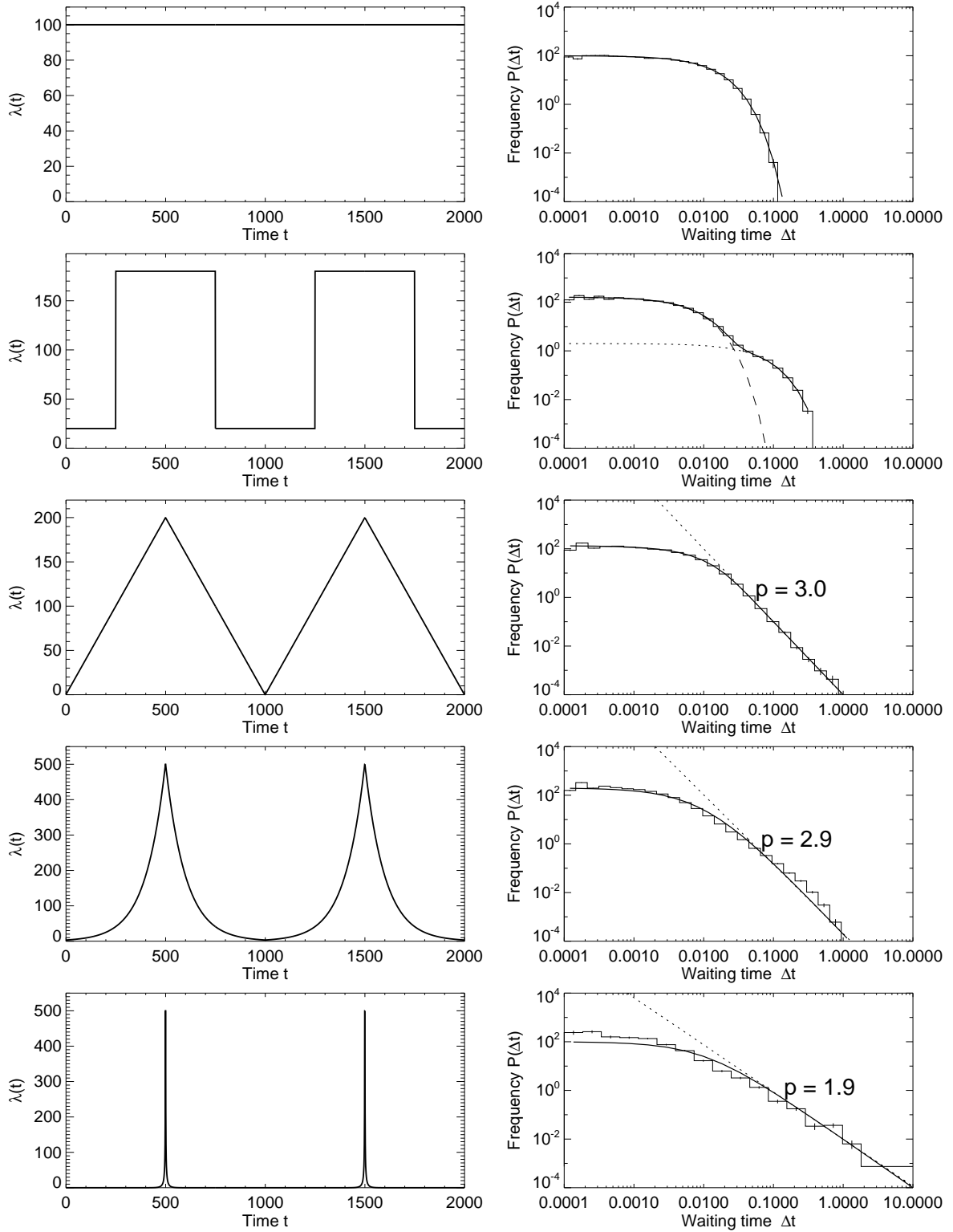}
\caption{One case of a stationary Poisson process (top) and four cases of 
nonstationary Poisson processes with two-step, linear-increasing, 
exponentially varying, and $\delta$-function variations of the occurrence rate 
$\lambda(t)$. The time-dependent occurrence rates 
$\lambda(t)$ are shown on the left side, while the waiting time distributions 
are shown in the right-hand panels, in form of histograms 
sampled from Monte-Carlo simulations, as well as in form of the analytical
solutions (given in Eqs.~10-18). Powerlaw fits $N(\Delta t)
\propto \Delta t^{-p}$ are indicated with
a dotted line and labeled with the slope $p$.}
\end{figure}

\begin{figure}
\plotone{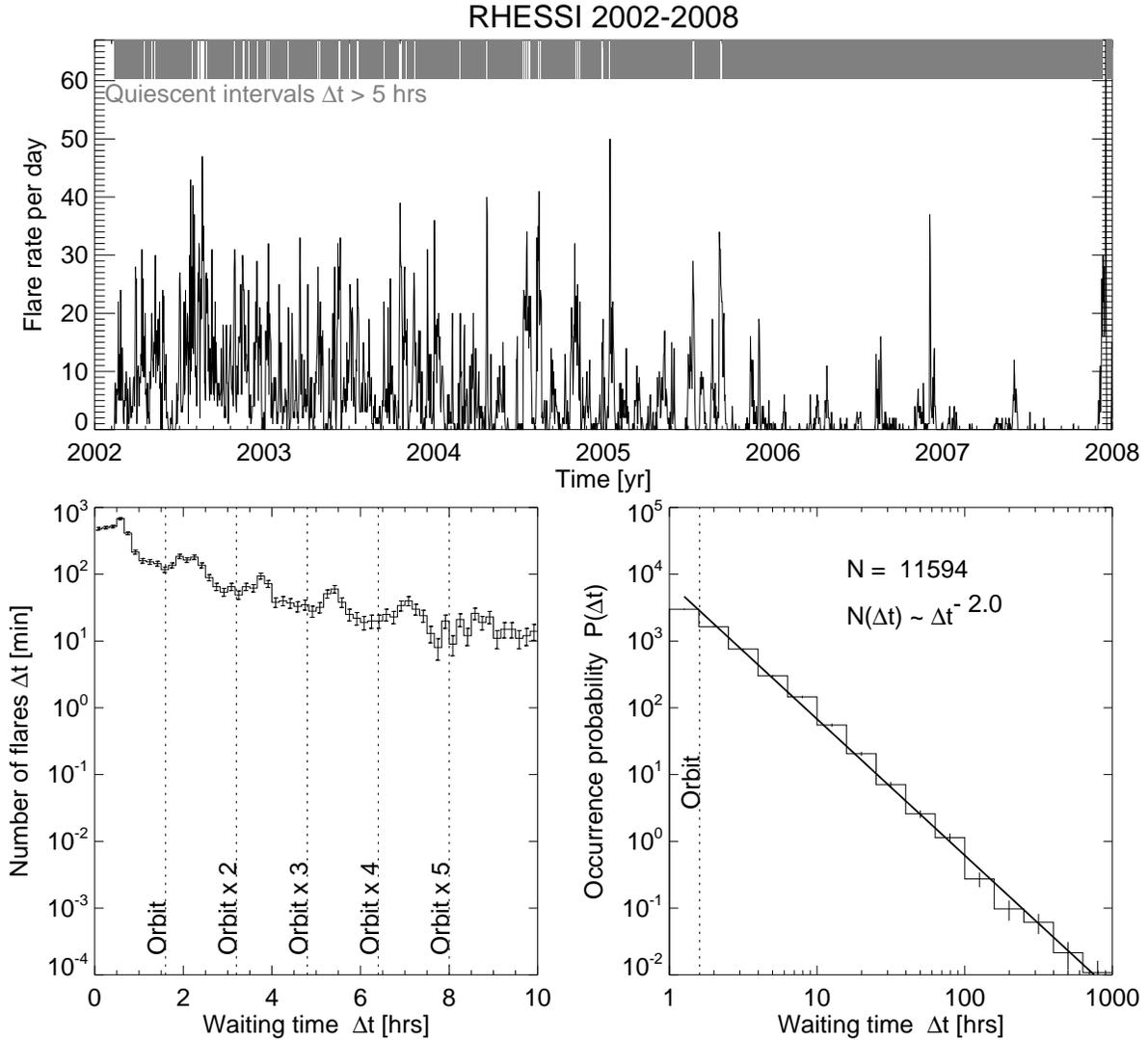}
\caption{{\sl Top:} Flare rate per day observed with RHESSI during
2002-2009, containing a total of 52,014 events. 
Quiescent time intervals with $\Delta t > 5$ hrs are marked in the form of
a ``bar code'' at the top of the panel. 
{\sl Bottom left:} The frequency distribution
of waiting times $N(\Delta t)$ is shown for short time intervals
$\Delta t \le 10$ hrs, which shows peaks at subharmonics of the
orbital period of $\approx 1.6$ hrs. 
{\sl Bottom right:} The longer waiting time intervals
$\Delta t \approx 2-24$ hrs can be fitted with a powerlaw function
with a slope of $p = 2.0$.}
\end{figure}

\begin{figure}
\plotone{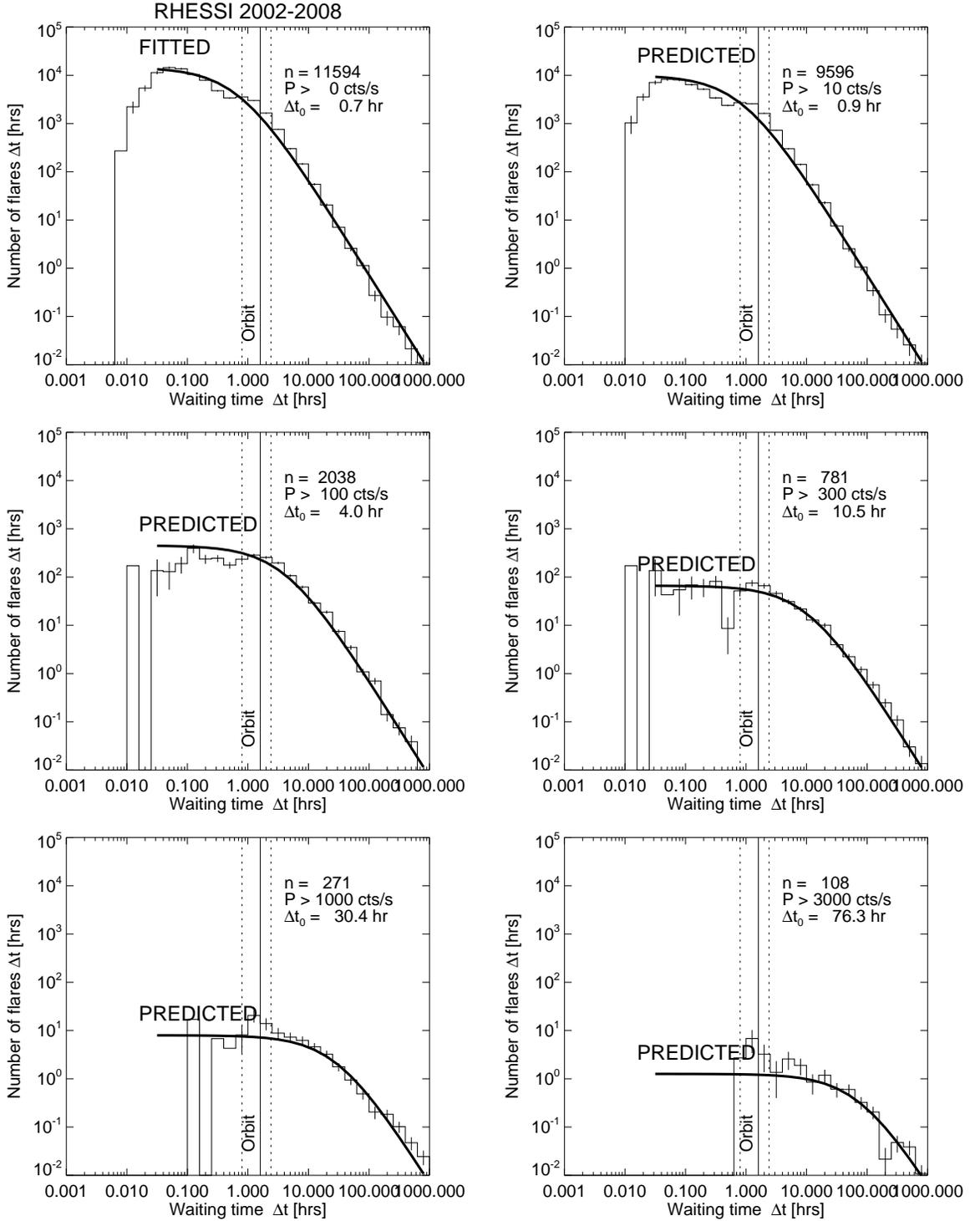}
\caption{Waiting time distributions for six different subsets of the data
selected by thresholds, $P \ge 0, 10, 100, 300, 1000, 3000$ cts s$^{-1}$.
The full distribution (with no threshold) is fitted with a model
of a nonstationary Poisson process (Eq.~18) with a powerlaw slope of $=2$ 
(thick solid curve in top left panel). The same model functions are predicted 
(using the scaling of Eq.~21) for the threshold-selected five subsets based 
on the number of detected events above the threshold.}
\end{figure}

\begin{figure}
\plotone{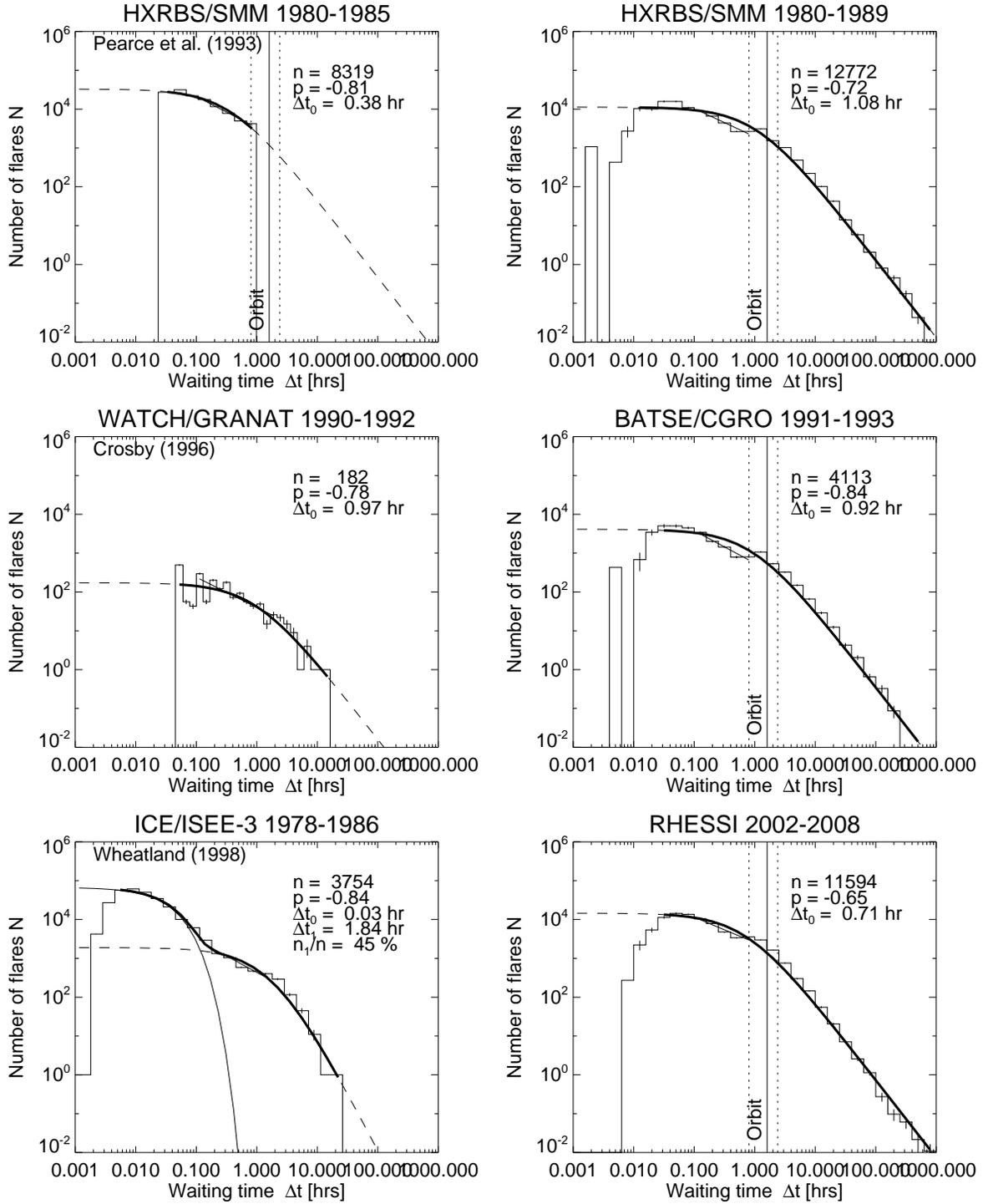}
\caption{Waiting time distributions of six different datasets:
HXRBS/SMM (top left and right), WATCH/GRANAT (middle left), 
ISEE-3/ICE (bottom left), BATSE/CGRO (middle right), and
RHESSI (bottom right). The distribution of the observed waiting times
are shown with histograms, fits of nonstationary Poisson processes with 
dashed curves (with a powerlaw tail with slope of $p=2$), and the best fit 
in the fitted range with thick solid curves. 
Powerlaw fits in the range of $\Delta t \approx 0.1-2.0$ hrs as fitted in the 
original publications (Pearce et al.~1993; Crosby 1996) are also shown 
(straight line and slope $p$).
The excess of events with waiting times near the orbital period 
($t_{orbit} \approx 1.6$ hrs) is an artificial effect and is not 
included in the model fits.}
\end{figure}

\end{document}